\documentclass[prb, superscriptaddress, preprint, showkeys]{revtex4}
\usepackage{graphicx}
\usepackage{amsmath}
\usepackage{array}
\newcolumntype{L}[1]{>{\raggedright\let\newline\\\arraybackslash\hspace{0pt}}m{#1}}
\newcolumntype{C}[1]{>{\centering\let\newline\\\arraybackslash\hspace{0pt}}m{#1}}
\newcolumntype{R}[1]{>{\raggedleft\let\newline\\\arraybackslash\hspace{0pt}}m{#1}}
\usepackage{amssymb}
\usepackage[T1]{fontenc}
\usepackage[english]{babel}
\usepackage{mathrsfs}
\usepackage{hhline}
\begin{document}
\author{Kirsten Govaerts}
\address{EMAT, Department of Physics, Universiteit Antwerpen, Groenenborgerlaan 171, 2020 Antwerpen, Belgium}
\author{Bart Partoens}
\address{CMT group, Department of Physics, Universiteit Antwerpen, Groenenborgerlaan 171, 2020 Antwerpen, Belgium}
\author{Dirk Lamoen}
\email{Dirk.Lamoen@uantwerpen.be}
\address{EMAT, Department of Physics, Universiteit Antwerpen, Groenenborgerlaan 171, 2020 Antwerpen, Belgium}
\title{Extended homologous series of Sn-O layered systems: a first-principles study}

\begin{abstract}
Apart from the most studied tin-oxide compounds, SnO and SnO$_2$, intermediate states have been claimed to exist for more than a hundred years.
In addition to the known homologous series (Seko et al., Phys. Rev. Lett. 100, 045702 (2008)), we here predict the existence of several new compounds with an O
concentration between 50 \% (SnO) and 67 \% (SnO$_2$). All these intermediate compounds are constructed from removing one or more (101) oxygen layers of SnO$_2$.
Since the van der Waals (vdW) interaction is known to be important for the Sn-Sn interlayer distances, we use a vdW-corrected functional, and compare these results with
results obtained with PBE and hybrid functionals. We present the electronic properties of the intermediate structures and we observe a decrease of the band gap when (i) the O
concentration increases and (ii) more SnO-like units are present for a given concentration. The contribution of the different atoms to the valence and conduction band is also
investigated.
\end{abstract}


\maketitle

\section{Introduction}

SnO$_2$ and SnO are the most studied tin-oxide compounds, showing a wide variety of technological applications. The former is most known as an opacifier of glazes \cite{sno_glazer},
it can also be used as a polishing powder \cite{sno_polish}, it can help to adhere a protective polymer coating \cite{sno_coat1} and it acts as a functional material for solar cells
\cite{sno_solar}, transparent conducting oxides \cite{sno_tco,sno_tco2} and gas sensors \cite{sno_sensor}. SnO is used for the production of tin salts for electroplating \cite{sno_coat2}
 and is known as an anode material for Li-rechargeable batteries \cite{sno_batt}. Moreover SnO is a good p-type semiconductor \cite{exp2}, which can even be converted from p- to
  n-type after doping with Sb \cite{hosono}.

\begin{figure}[!h]
    \centering
        \includegraphics[width=0.48\textwidth]{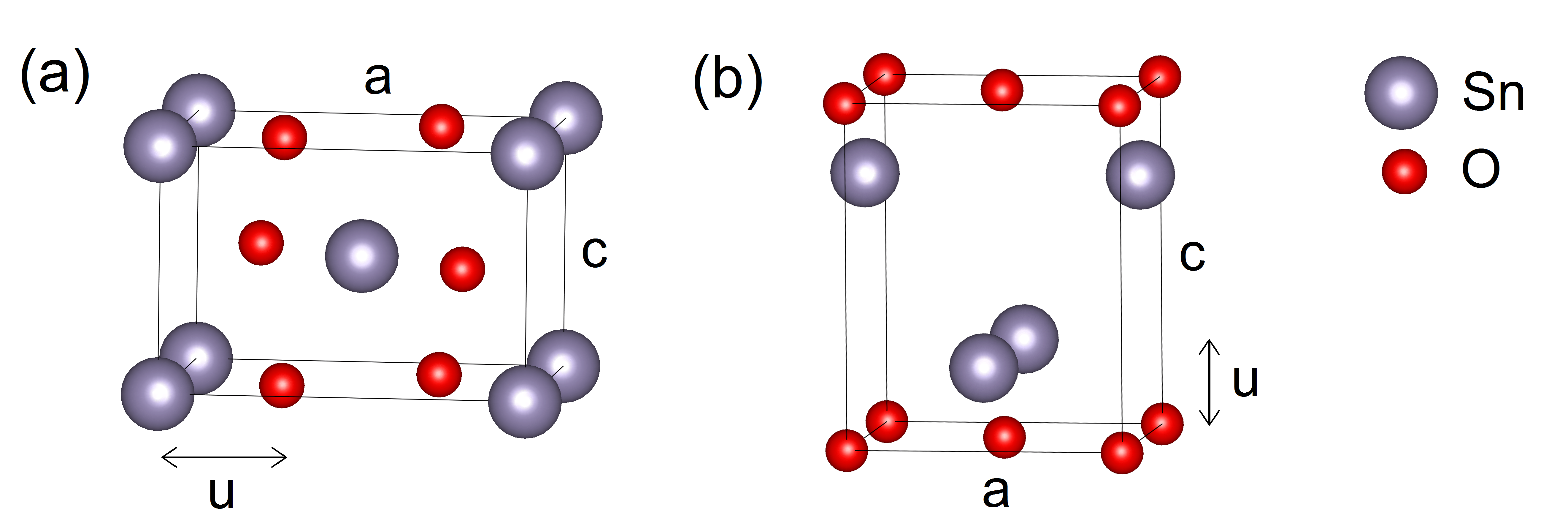}
    \caption{Unit cells of (a) SnO$_2$ and (b) SnO.}
    \label{fig:SnO2-SnO}
\end{figure}

Both SnO$_2$ and SnO have a tetragonal structure in their most
stable form. For SnO$_2$ this is the rutile structure (space group
$P4_2/mnm$) with lattice parameters $a = 4.7374$ \AA\: and $c =
3.1864$ \AA. In Fig. \ref{fig:SnO2-SnO} (a) it can be seen that
SnO$_2$ has two formula units per unit cell and an internal
parameter, $u = 0.30562$, fixing the positions of the O atoms
\cite{sno2_str}. The most stable form of SnO is the so-called
litharge structure (space group $P4/nmm$) with lattice parameters
$a = 3.8011$ \AA, $c = 4.8352$ \AA. In Fig. \ref{fig:SnO2-SnO} (b)
it can be seen that also SnO has two formula units per unit cell
with internal parameter,
 $u = 0.23818$ \cite{vdw2}, fixing the positions of the Sn atoms located at the apex of the square pyramid with a base of O atoms. This results in a layered stacking of Sn-O-Sn slabs
 along the [001] direction with each slab consisting of an oxygen layer sandwiched between two tin layers. Between the Sn-O-Sn slabs i.e. between adjacent Sn layers a weak van der
  Waals (vdW) bonding exists with a distance of 2.53 \AA\: between successive slabs.

Apart from these well-known compounds, intermediate states have been claimed to exist for more than a hundred years. Many phase diagrams have already been proposed in literature,
but they do not agree with each other \cite{cahen}. The intermediate states arise in the disproportionation of SnO at higher temperatures: apart from the (usually wanted) end
products SnO$_2$ and Sn, also (usually unwanted) intermediate oxides occur. Additionally, SnO$_2$ often shows oxygen deficiency, which plays a central role in determining properties
and chemical activities. Recently it was shown that a Sn$_2$O$_3$-based NO gas sensor demonstrated an improved selectivity to NO, NO$_2$, and CO gases \cite{gas}. Another study
remarks the presence of Sn$_3$O$_4$ in ultrathin SnO$_2$ nanosheets, which show superior performance for lithium-ion storage \cite{sno2-nanosheets}. The compounds Sn$_3$O$_4$
\cite{spandau,spinedi,ditte,berengue,holleman,lawson,moreno,cahen}, Sn$_2$O$_3$ \cite{klushin,murken} and to a lesser extent Sn$_5$O$_6$ \cite{decroly}, have already been the subject
of research.

\begin{figure}[!h]
    \centering
        \includegraphics[width=0.48\textwidth]{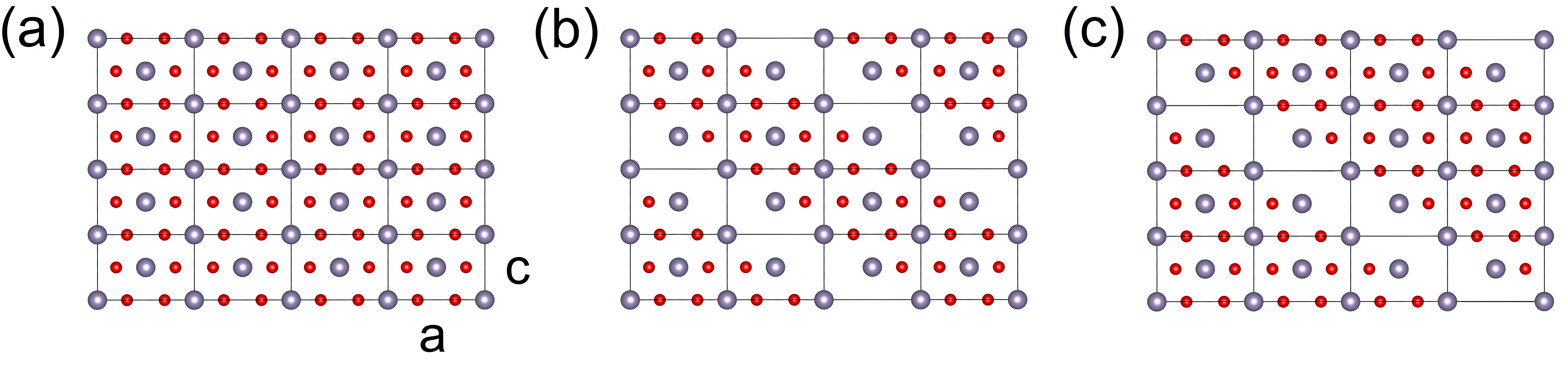}
    \caption{Crystal structures of (a) SnO$_2$, (b) Sn$_3$O$_4$ and (c) Sn$_2$O$_3$.}
    \label{fig:SnO2-Sn3O4-Sn2O3}
\end{figure}

Based on a cluster expansion (CE) in combination with simulated annealing, Seko et al. \cite{seko} suggested the existence of a homologous series Sn$_{n+1}$O$_{2n}$ of (meta)stable
compounds. These structures have oxygen vacancies layered on (101) planes of the rutile lattice of SnO$_2$, as shown in Fig. \ref{fig:SnO2-Sn3O4-Sn2O3} for Sn$_3$O$_4$ and Sn$_2$O$_3$
structures. For the latter, this was already observed in Ref. \onlinecite{sn2o3} where different possible structures were investigated with ab initio techniques, resulting in a lowest
energy for the structure with oxygen vacancies layered on the (101) planes of SnO$_2$.

Data concerning the stability and the details of the crystal structures of the intermediate compounds, however, is very rare. The foregoing results have motivated us to search
for new Sn-O compounds and to investigate their stability and electronic properties.

\begin{figure}[!h]
    \centering
        \includegraphics[width=0.5\textwidth]{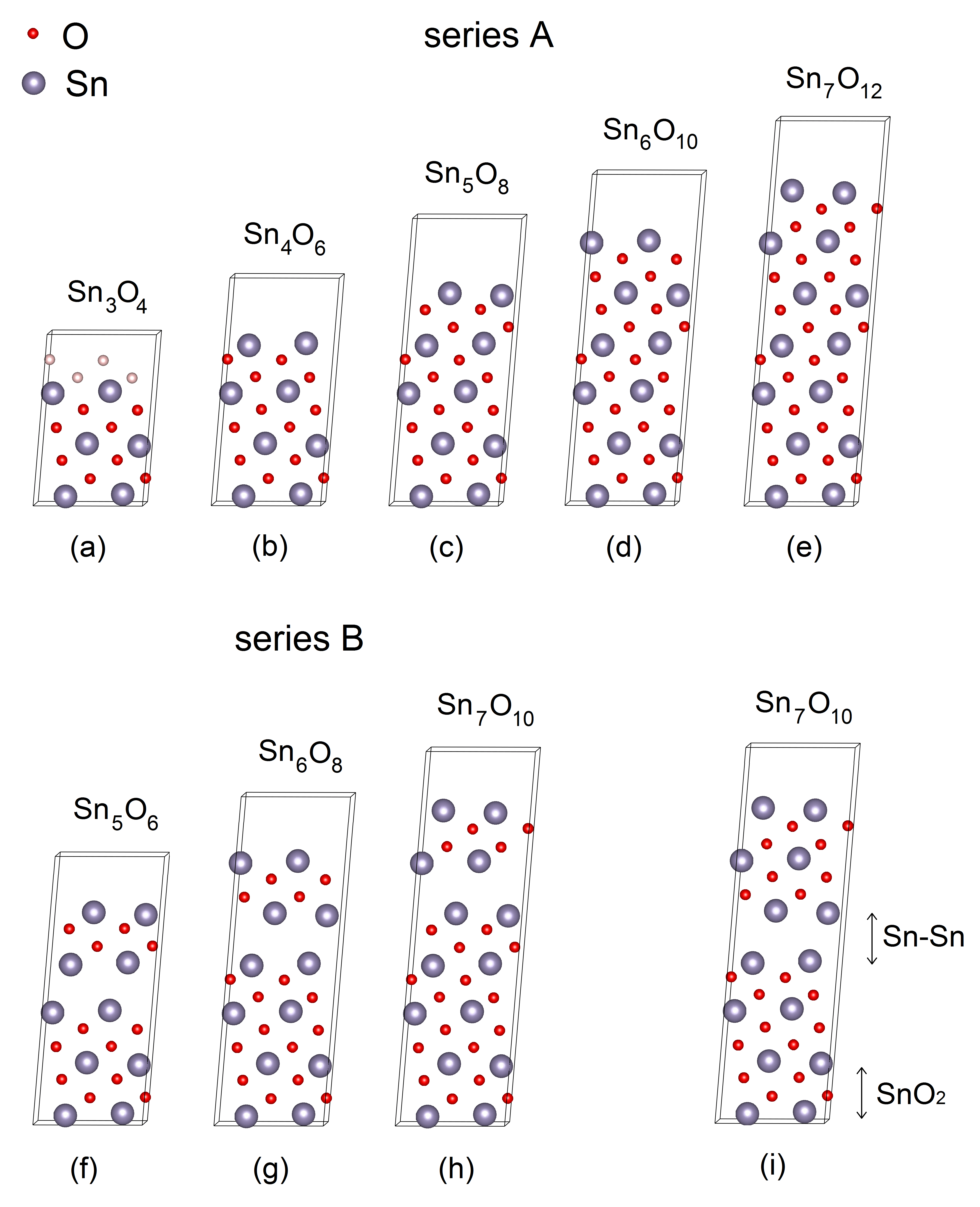}
    \caption{The unit cells of nine different structures, resulting from removing one or more oxygen layers on (101) planes from the SnO$_2$ structure. In structure (a) the removed
    oxygen layer is shown by the light atoms. The structures of series A - structures (a)-(e) - are elements of the homologous series Sn$_{n+1}$O$_{2n}$ with $n = 2,3,4,5,6$
    respectively. Structures (f)-(i) are new, where structures (f)-(h) - series B - exhibit a small unit of an oxygen layer in between two Sn layers (i.e. SnO unit), structure (i)
    is composed of two structures of series A. Each unit cell consists of two formula units. In structure (i), the Sn-Sn interlayer distance and the thickness of the SnO$_2$ unit
    are denoted by arrows.}
    \label{fig:str-sno}
\end{figure}

In this study we investigate first the stability of different proposed Sn-O structures from the known homologous series Sn$_{n+1}$O$_{2n}$ with $n = 2,3,4,5,6$. When relaxing
these structures, we start from a SnO$_2$ superstructure and remove one oxygen layer, resulting in unit cells consisting of two formula units. This can be seen in the upper
part of Fig. \ref{fig:str-sno}, where in structure (a), the removed O layer is shown by the light atoms. In the following we will refer to these structures as series A. In
addition to this homologous series Sn$_{n+1}$O$_{2n}$ we also removed two non-successive O layers, resulting in the structures shown in the lower part of Fig. \ref{fig:str-sno}.
For structures (f)-(h), this results in a small unit with only one O layer in between two Sn layers - further referred to as SnO unit - and a larger building block of series A,
(a)-(c) respectively. We will refer to these structures as series B. In structure (i) we also removed two O layers, now resulting in a unit cell consisting of two building blocks
of series A, (a) and (b). The structures (f)-(i) are not mentioned by Seko et al. but can also be described by a homologous series Sn$_{n+2}$O$_{2n}$ and the unit cells again consist
of two formula units. We investigate their stability and compare this with the stability of the homologous series Sn$_{n+1}$O$_{2n}$, i.e. series A.

In Ref. \onlinecite{sno} it was shown that the vdW interaction is important between two adjacent Sn-layers. Therefore we compare the stability of the intermediate Sn-O structures
obtained by relaxing them with and without the use of a vdW-corrected functional. In a second part of this study the structural properties of the structures are investigated using
different exchange- and correlation (xc) functionals: a generalized gradient approximation, a vdW-corrected density functional and a hybrid functional. We relax both the lattice
parameters and the atomic positions with the three functionals. Furthermore we discuss the electronic properties of these structures. Whereas the structural properties can be
understood from the extreme cases of SnO and SnO$_2$, the resulting electronic properties are not straightforward.

\section{Computational details}
\label{sec:twee} The Vienna \emph{ab initio} simulation package
VASP \cite{vasp1, vasp2} was used to optimize lattice parameters
and atomic positions. We used the all-electron projector augmented
wave (PAW) method with the Sn ($5s^2 5p^2 4d^{10}$) and O ($2s^2
2p^4$) electrons treated as valence electrons. For the xc
functional we considered both the generalized gradient
approximation of Perdew-Burke-Ernzerhof (PBE) \cite{PBE}, the
vdW-DF, optB86b, discussed in ref.\cite{optB}, and the hybrid
functional proposed by Heyd, Scuseria and Ernzerhof (HSE06) using
a mixing parameter $\alpha = 0.25$ \cite{hf1,hf5}.

For total energy calculations and structure optimization we used a
$\Gamma$-centered Monkhorst-Pack $k$-points grid \cite{MP} for the
Brillouin zone integration. For the PBE and vdW calculations, the
plane wave cutoff value was 800 eV, and for the HSE 600 eV. The
total number of k-points was chosen so that our results are
converged within 10$^{-4}$ eV/atom. The results were considered
converged when the energy difference between two successive steps
was smaller than 10$^{-5}$ eV and for the geometry optimization we
considered a convergence criterium for the forces on the atoms of
less than 10$^{-3}$ eV/\AA\: for the PBE and vdW calculations, and
5 $\cdot$ 10$^{-3}$ eV/\AA\: for the HSE calculation.

Only for SnO and SnO$_2$ we performed $G_0W_0$ calculations
\cite{shishkin,shishkin1}, where we used a 6$\times$6$\times$6 and
4$\times$4$\times$6 $k$-point grid respectively, 80 empty bands
and a cutoff energy of 600 eV.

\section{Results}

\subsection{Energetic stability}

The nine structures of Fig. \ref{fig:str-sno} are the compounds with the smallest unit cells with a composition between SnO and SnO$_2$, that can be formed when removing one or two
O layers from a SnO$_2$ structure. In Ref. \onlinecite{sno} it was shown that the vdW interaction plays an important role in the SnO structure, particularly between adjacent Sn layers.
In the intermediate Sn-O structures, the O layers between two Sn layers are removed, resulting in adjacent Sn layers, where we expect the vdW interaction to be non negligible.

\begin{figure}[!h]
    \centering
        \includegraphics[width=0.5\textwidth]{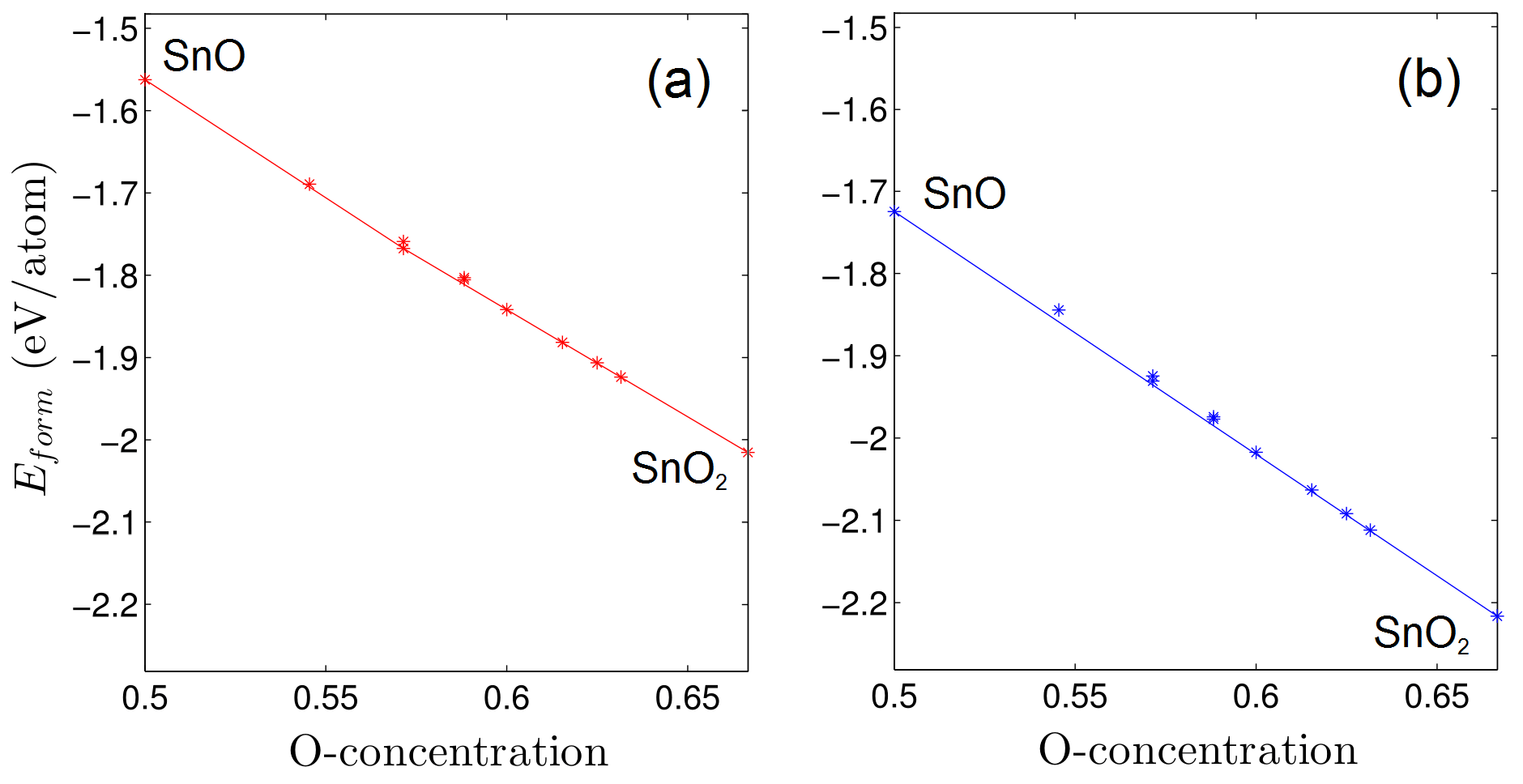}
    \caption{Convex hull of Sn-O structures relaxed with (a) PBE and (b) optB86b-vdW.}
    \label{fig:hulls}
\end{figure}

To investigate this effect, we relax the structures with the regular PBE functional and with the optB86b-vdW functional. Formation energies - with respect to O$_2$ and elemental Sn -
of all structures with these two functionals are shown in Fig. \ref{fig:hulls} (a) and (b) respectively, for O concentration ranging from 50 \% (i.e. SnO) to 66 \% (i.e. SnO$_2$). For
the PBE-relaxed structures the convex hull of these formation energies is similar to the one of Seko et al \cite{seko}. Structures of series A are found to be ground state structures.
The convex hull of the vdW-relaxed structures however, shows that these structures are all slightly above the convex hull (2-5 meV/atom), and should therefore be considered as
metastable structures. However, in principle entropic effects can stabilize these structures at finite temperature and in any case the energy difference of these metastable structures
 with the convex hull is small enough to be observed experimentally \cite{ceder2}. One should also keep in mind that vdW-DFs have been tested less in comparison with the standard PBE
 functional when it comes to the accurate prediction of ground state energies. We also relaxed the structures using a hybrid functional, and their stability is similar to the PBE-relaxed
  structures.

In addition to the homologous series Sn$_{n+1}$O$_{2n}$ (series A) we considered structures from the homologous series Sn$_{n+2}$O$_{2n}$ (series B and structure (i) of Fig.
\ref{fig:str-sno}). These structures are metastable at T = 0 K because their formation energy is slightly above the convex hull (2-7 meV/atom), both for the PBE- and vdW-relaxed
structures. Again we would like to emphasize that such small energy differences could very well be compensated at elevated temperatures by entropic effects \cite{ceder2}. Structures
of series B are composed of a larger building block of series A, and one small SnO unit, while structure (i) is composed of two building blocks of series A. Structures (h) and (i)
have the same composition, but O layers are removed at different places, resulting in structures composed of different building blocks. From the formation energies of these structures
we conclude that structure (i) is favored by 2.6 meV/atom. This means that the Sn-O compounds favor the formation of larger building blocks.

If we remove more O layers resulting in a structure with 50 \% O, it did not relax to the SnO ground state structure. It is trapped in a local minimum, which differs by 6.3 meV/atom
from the stable litharge structure of SnO. In the following, when SnO is mentioned, one should keep in mind that we refer to its stable litharge ground state structure.

\subsection{Structural properties}

When discussing the structural properties of the intermediate Sn-O structures, we compare the two most relevant features: the Sn-Sn interlayer distance and the thickness of
the SnO$_2$ unit, from which the structures are built up. These two features are shown in Fig. \ref{fig:str-sno} (i). Besides using PBE and optB86b-vdW functionals, we also relaxed
the structures using the HSE06 hybrid functional.

\begin{figure}[!h]
    \centering
        \includegraphics[width=0.48\textwidth]{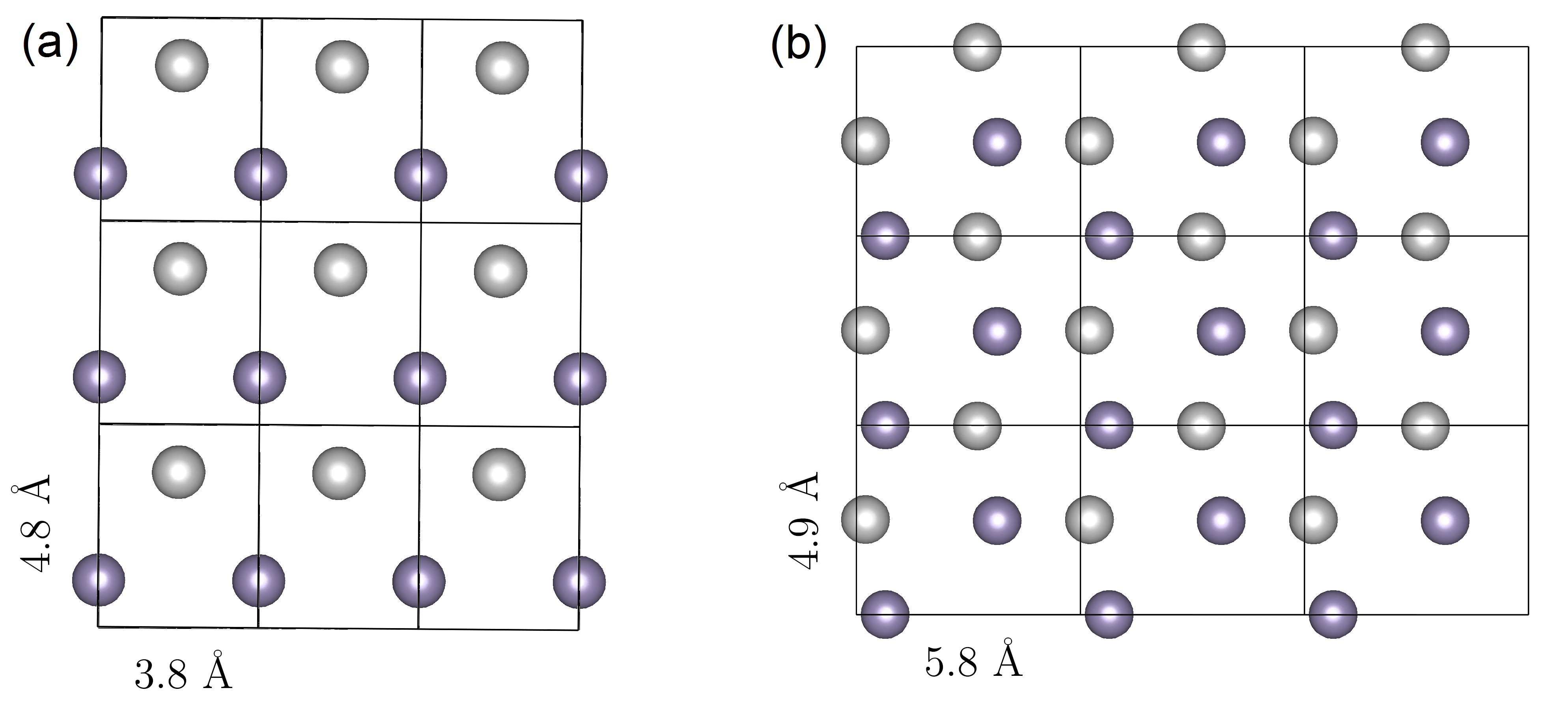}
    \caption{Top view of the two adjacent Sn layers (light and dark atoms) for (a) SnO and (b) Sn$_3$O$_4$.}
    \label{fig:sn-layers}
\end{figure}

In Ref. \onlinecite{sno}, the Sn-Sn interlayer distance of SnO was investigated. As mentioned before, the SnO ground state structure can not be obtained by removing O layers from
the SnO$_2$ structure. In Fig. \ref{fig:sn-layers} it is shown that the ordering of Sn atoms in the layers of SnO and the intermediate structures is clearly different. Fig.
\ref{fig:sn-layers} (a) shows a top view of the adjacent Sn-layers of SnO, while Fig. \ref{fig:sn-layers} (b) shows the top view of the adjacent Sn-layers of a typical member
of the homologous series, Sn$_3$O$_4$. Since this ordering of Sn atoms in the layers is different, interlayer distances will be different.

For the Sn-Sn interlayer distance of SnO, the optB86b-vdW functional gives results which are in very good agreement with experiment, as shown in Table \ref{table:sno-layers}. PBE
and HSE overestimate the interlayer distance, although HSE is somewhat better.

Similar results are obtained for the intermediate structures, of which the Sn-Sn interlayer distances are shown in Table \ref{table:sno-layers}. When comparing the results for the
different functionals, the analogy to SnO is clear: PBE leads to the largest values, while the hybrid functional leads to smaller ones. vdW yields the smallest Sn-Sn interlayer
distances.
We also observe smaller Sn-Sn interlayer distances for the structures of series B. Structure (i), where two O layers are removed resulting in a structure consisting of building blocks
of Sn$_3$O$_4$ and S$_4$O$_{6}$ has a Sn-Sn interlayer distance in between that of of Sn$_3$O$_4$ and Sn$_4$O$_{6}$.

\begin{table}[!htb]
\centering
\begin{tabular}{|L{1.7cm} || C{1.5cm} | C{1.5cm} | C{1.5cm} | C{1.0cm} |}
\hline
& PBE & vdW & hybrid & exp \\ \hhline{|=||=|=|=|=|}
SnO & 2.705 & 2.492 & 2.640 & 2.533 \\ \hline
(a)-(e)+(i) & 3.05-3.1 & 2.81-2.82 & 2.97-3.00 & / \\ \hline
(f)-(h) & 3.02-3.05 & 2.78-2.79 & 2.94-2.97 & /\\ \hline
\end{tabular}
\caption{Sn-Sn interlayer distances of SnO and the different structures of Fig. \ref{fig:str-sno} (in \AA).}
\label{table:sno-layers}
\end{table}

We can also compare the thickness of the SnO$_2$ unit. In Table \ref{table:sno2-layers} this thickness is shown for SnO$_2$, relaxed with different functionals, and compared with the
experimental value. The different values do not differ as much as in the case of SnO, and are all pretty close to the experimental value. All functionals slightly overestimate the
SnO$_2$ thickness, but the value obtained with the hybrid functional is closest to the experimental value.

The same trend can also be seen in the Sn-O structures of Fig. \ref{fig:str-sno}, for which the results are shown in Table \ref{table:sno2-layers}. The hybrid functional gives smaller
values for the thickness of the SnO$_2$ unit than the vdW and PBE functionals. The values obtained with the vdW functional are again almost similar for all structures, while the PBE
functional gives values of 2.65 \AA\: for the structure with the smallest concentration of O, and 2.69 \AA\: for the structure with the largest O concentration. In particular, when
the O concentration is smaller than 60 \%, the vdW-value is larger than the PBE-value, and when the O concentration exceeds 60 \%, the PBE-value gets larger. Also the values obtained
with the hybrid functional increase with increasing O concentration. Finally we remark that the thicknesses of SnO$_2$ units of structures (f)-(h) are comparable with those of their
corresponding SnO$_2$ units of structures (a)-(c) respectively.

\begin{table}[!htb]
\centering
\begin{tabular}{|L{1.7cm} || C{1.5cm} | C{1.5cm} | C{1.5cm} | C{1.0cm} |}
\hline
& PBE & vdW & hybrid & exp \\ \hhline{|=||=|=|=|=|}
SnO$_2$ & 2.693 & 2.681 & 2.651 & 2.643 \\ \hline
(a)-(i) & 2.65-2.69 & 2.67-2.68 & 2.63-2.65 & / \\ \hline
\end{tabular}
\caption{Thickness of SnO$_2$ building block of SnO$_2$ and the different structures of Fig. \ref{fig:str-sno} (in \AA).}
\label{table:sno2-layers}
\end{table}

\subsection{Electronic properties}

Before looking at the band gap of the intermediate structures, we
present the results for the two known compounds SnO and SnO$_2$ in
Table \ref{table:sno-sno2-gaps}. The band gap calculated with PBE,
vdW and hybrid functionals belong to the structures relaxed with
these respective functionals. The band gap of the structure
relaxed with the vdW functional, is also calculated within the
$G_0W_0$ approximation.

\begin{table}[!htb]
\centering
\begin{tabular}{| L{1.3cm} || C{1.1cm} | C{1.1cm} | C{1.1cm} | C{1.1cm} | C{1.1cm} |}
\hline
& PBE & vdW & hybrid & $G_0W_0$ & exp \\ \hhline{|=||=|=|=|=|=|}
SnO & 0.43 & 0.16 & 0.59 & 0.60 & 0.7 \\ \hline
SnO$_2$ & 0.66 & 0.84 & 2.95 & 2.49 & 3.6 \\ \hline
\end{tabular}
\caption{Band gap (in eV) of SnO and SnO$_2$ for different functionals.}
\label{table:sno-sno2-gaps}
\end{table}

In the case of SnO, $G_0W_0$ gives a band gap in very good agreement with the experiment, however the value obtained with the hybrid functional is also close to the experimental value.
In the case of SnO$_2$ the hybrid functional clearly gives the best result. We should remark that the $G_0W_0$ results strongly depend on the starting wave functions, and here we used
starting wave functions from a PBE calculation.
 Since structural differences are small when SnO$_2$ is relaxed using the different functionals, the difference in band gap is mainly due to changes in the electron density as a result
 of the different functionals. This is not the case for SnO, where not only the use of a specific functional for calculating the band gap, but also the large differences in structural
  parameters - arising when relaxing with the different functionals - cause differences in the band gap.

\begin{figure}[!h]
    \centering
        \includegraphics[width=0.48\textwidth]{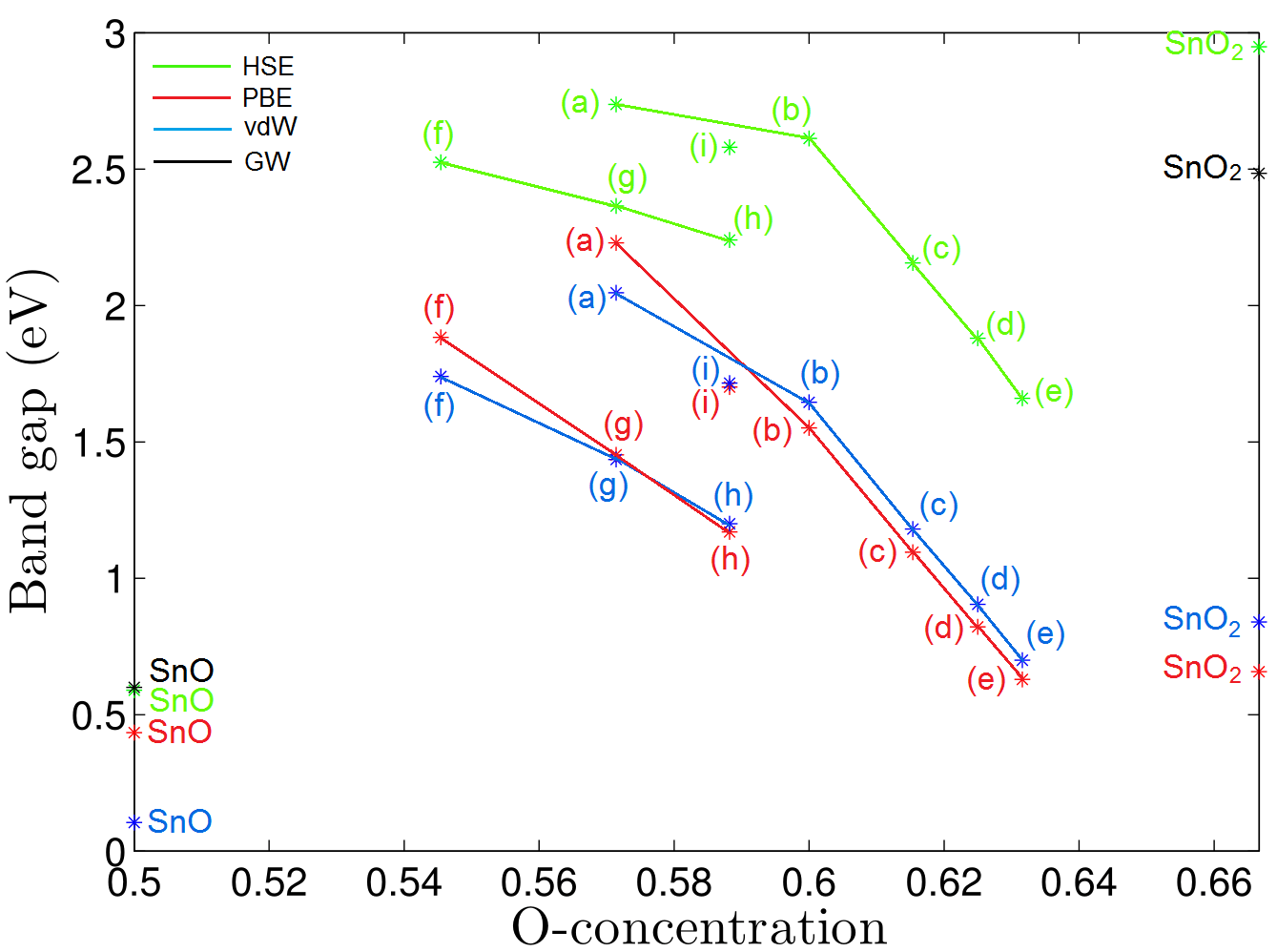}
    \caption{Band gap of all Sn-O structures of Fig. \ref{fig:str-sno}, calculated with the PBE (red), optB86b-vdW (blue) and HSE06 (green) functionals. SnO and SnO$_2$ band
    gaps are also calculated within the $G_0W_0$ approximation, on top of the vdW calculation (black).}
    \label{fig:allgaps}
\end{figure}

Whereas for SnO, the vdW gap is smaller than the PBE gap, the opposite is true for SnO$_2$. Therefore one expects a transition somewhere in between. Figure \ref{fig:allgaps} shows
the band gap for all structures, calculated with the different functionals. For the structures with an O concentration smaller than 58 \%, PBE gives indeed a larger band gap than
the vdW-DF. The vdW and PBE calculated band gap do not differ much, while the hybrid functional opens the gap with $\sim$ 1 eV. The band gap of the structures of series A monotonically
decrease with an increasing O concentration. The same trend can be observed for the band gap of the structures of series B. Structures (a) and (g) have the same concentration but
whereas in (a) only one O layer is removed from the initial SnO$_2$ superstructures, in (g) two O layers are removed, resulting in an additional SnO unit, which makes the band gap
decrease by 0.5-0.7 eV. The same decrease of the band gap can be observed for structures (h) and (i), which are both a result of removing two O layers from the SnO$_2$ structure,
but whereas for structure (h) this results in a building block of series A and a small SnO unit, for (i) it results in two building blocks of series A. Therefore we can conclude that
the decrease of the band gap is not a result of removing more O layers, but is mainly due to the existence of this small SnO unit in the structure.

Since the hybrid functional underestimates the band gap of SnO and SnO$_2$ by 15-17 \%, one can expect that the band gap of the intermediate compounds calculated with the hybrid
functional will also be underestimated by the same amount. There are no experimental (or theoretical) results available to compare with.

Although the band gap decreases monotonically with increasing O concentration, it increases again for SnO$_2$. In particular for the HSE calculations a steep increase is observed in
going from Sn$_{7}$O$_{12}$ (e) to SnO$_2$.

\begin{figure}[!h]
    \centering
        \includegraphics[width=0.48\textwidth]{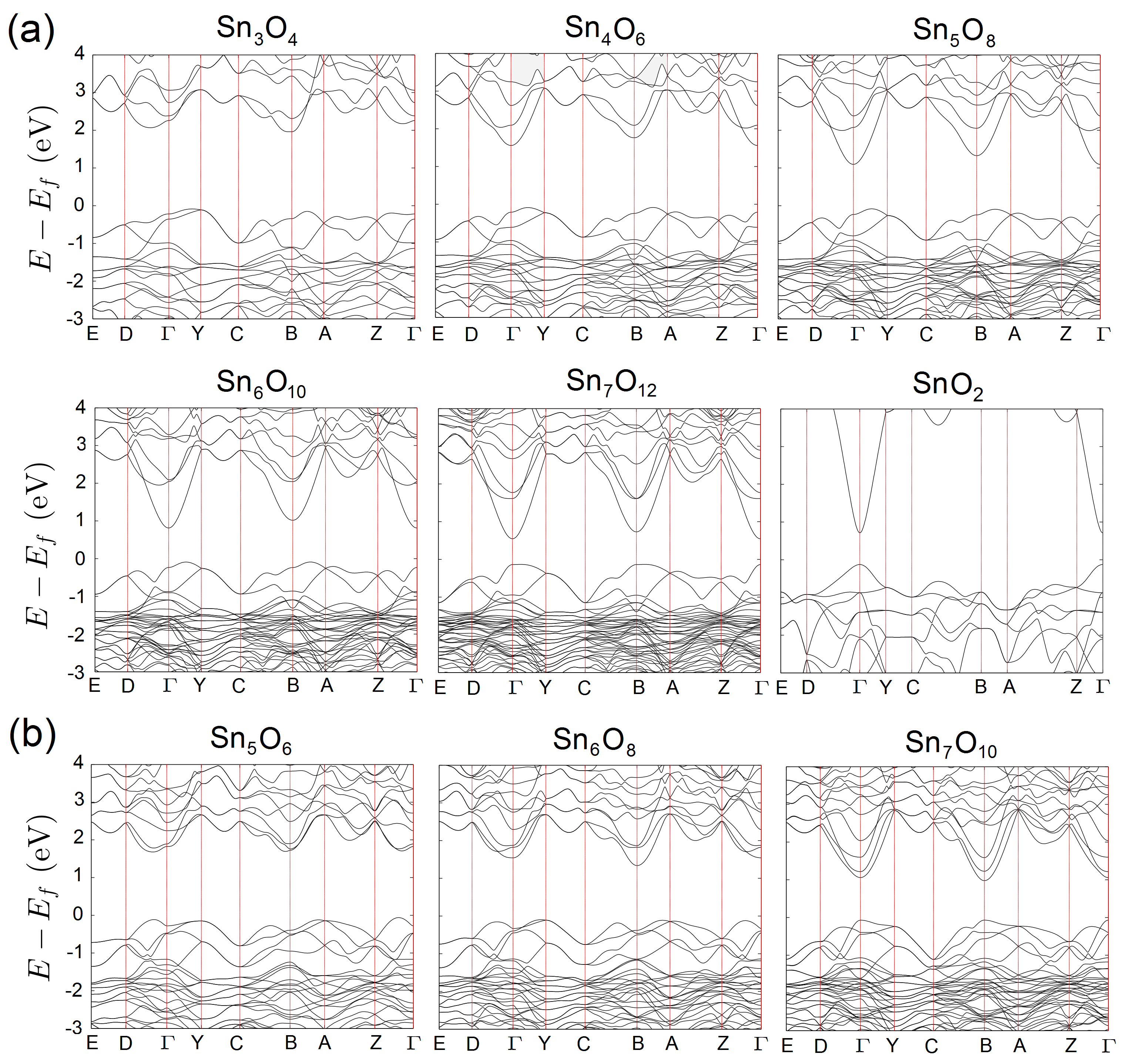}
    \caption{Band structure of structures of (a) series A, and in addition SnO$_2$, and (b) series B.}
    \label{fig:bs_all_sno}
\end{figure}

In Fig. \ref{fig:bs_all_sno} (a) and (b) the band structure of series A and series B, calculated with the vdW-DF, is shown respectively. For both figures, one can see the decrease
of the band gap with increasing O concentration.

\begin{figure}[!h]
    \centering
        \includegraphics[width=0.48\textwidth]{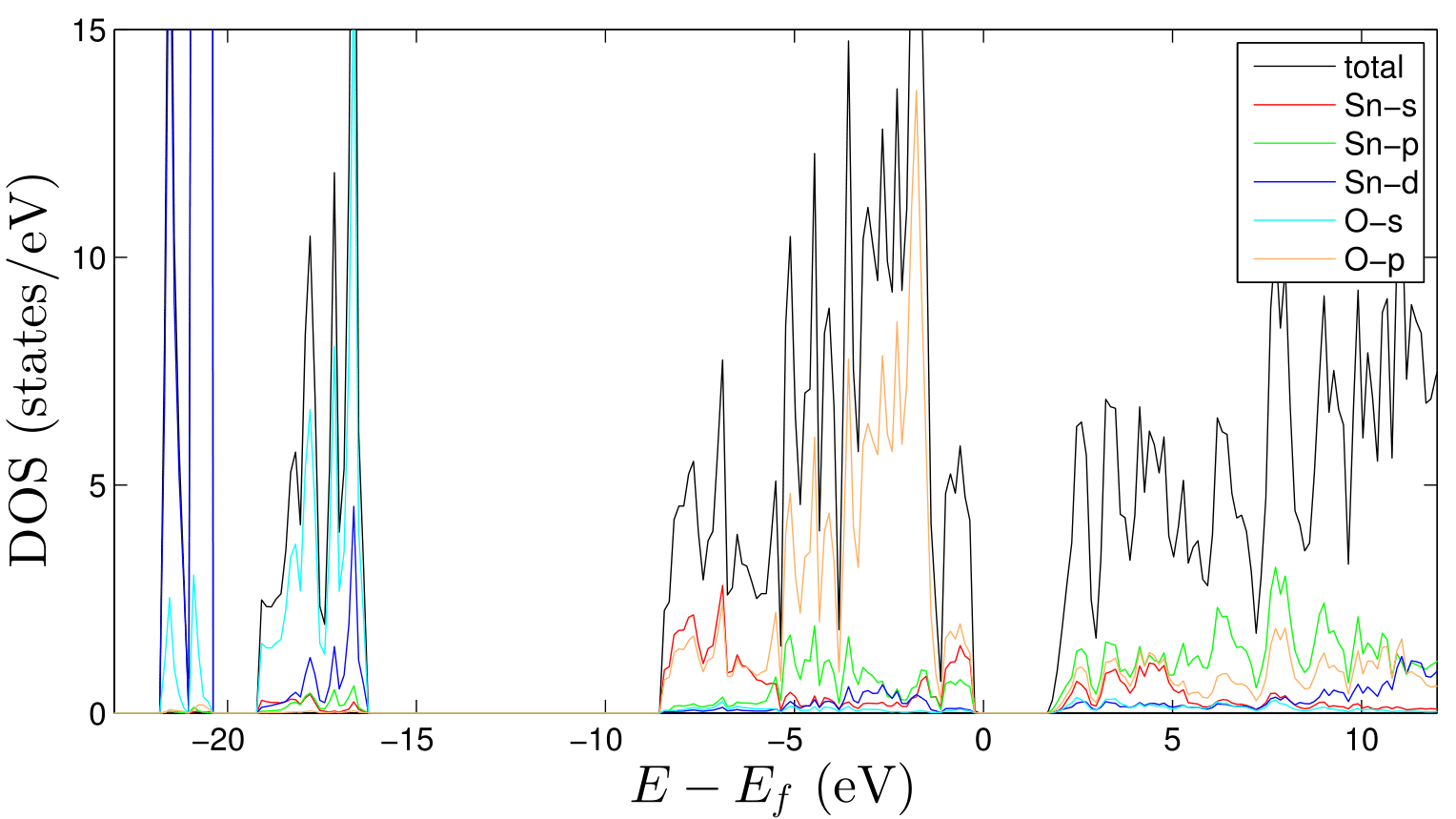}
    \caption{Density of states of Sn$_3$O$_4$. Black lines correspond to the total DOS, colored lines correspond to the projected DOS, as described in the legend.}
    \label{fig:DOS_Sn6O8}
\end{figure}

\begin{figure}[!h]
    \centering
        \includegraphics[width=0.24\textwidth]{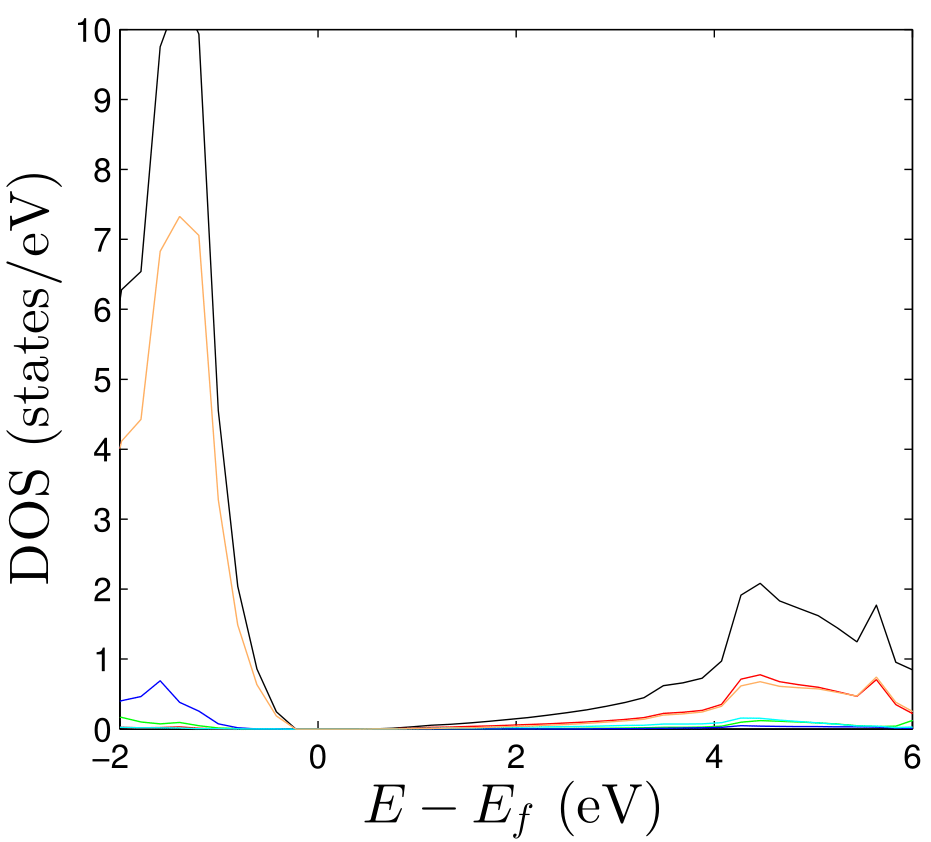}
    \caption{Density of states of SnO$_2$ around the band gap. The meaning of the colors is the same as in Fig. \ref{fig:DOS_Sn6O8}.}
    \label{fig:DOS_SnO2}
\end{figure}

To gain more insight in the previous findings and confirm earlier claims, we have investigated the contribution of the different atoms to the valence and conduction bands,
by calculating the projected DOS. In Fig. \ref{fig:DOS_Sn6O8} we show the DOS of Sn$_3$O$_4$ together with its PDOS. The top of the valence band consists of $\sim$ 47 \% O-p,
$\sim$ 35 \% Sn-s and $\sim$ 17 \% Sn-p states. These results are similar for the other intermediate structures and is in line with previous studies on the PDOS of SnO
\cite{sno,lonepair2}. The main contribution for the valence band of SnO$_2$ is O-p, as can be clearly seen in Fig. \ref{fig:DOS_SnO2}. These findings suggest that the valence
band of the intermediate structures is rather similar to the valence band of SnO.

\begin{figure}[!h]
    \centering
        \includegraphics[width=0.48\textwidth]{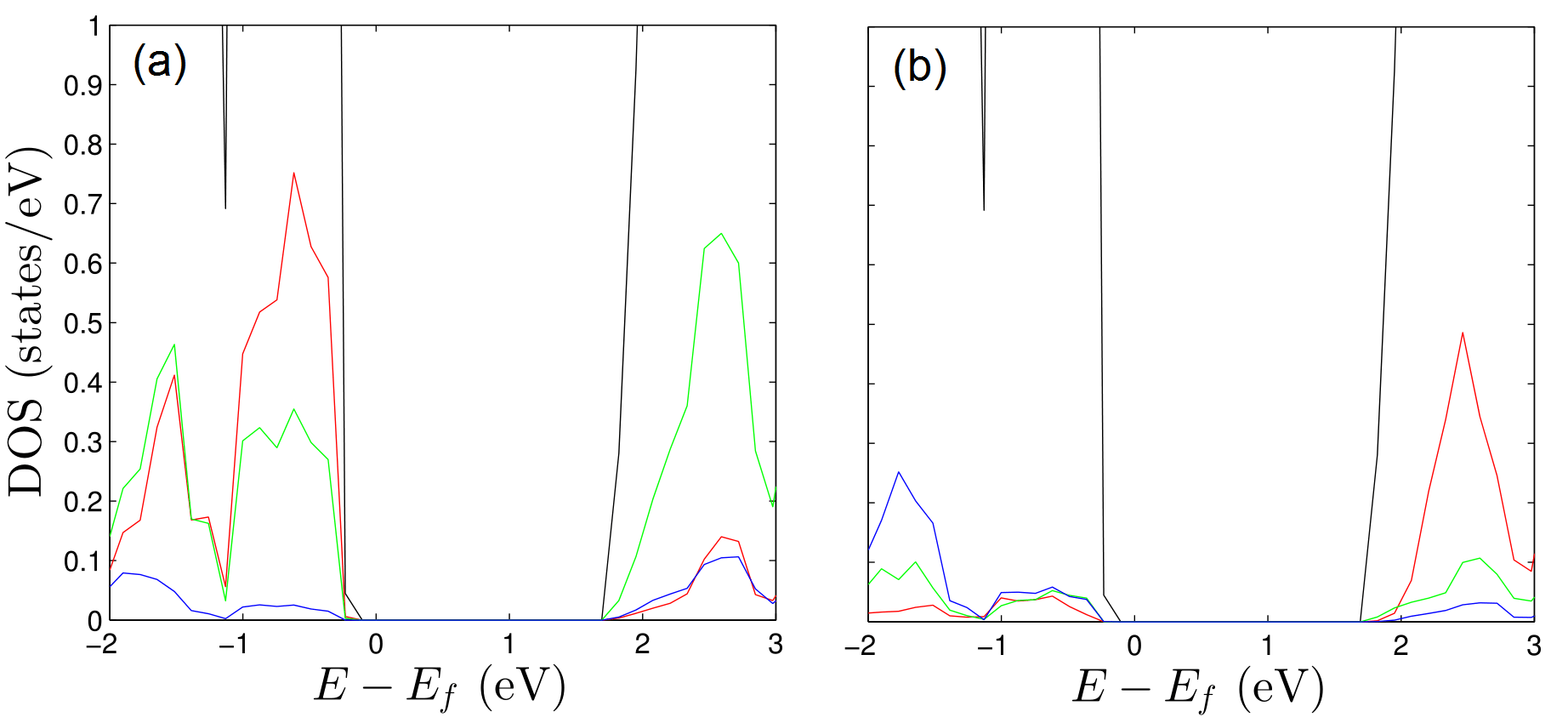}
    \caption{Density of states of Sn$_3$O$_4$ around the band gap. Projected DOS of the (a) SnO-like Sn atoms and (b) SnO$_2$-like Sn atoms. The meaning of the colors is the same
    as in Fig. \ref{fig:DOS_Sn6O8}.}
    \label{fig:DOS_Sn6O8_apart}
\end{figure}

In Fig. \ref{fig:DOS_Sn6O8_apart} (a) the contribution of the Sn atoms next to the removed O layer is shown, i.e. the SnO-like Sn atoms, and  Fig. \ref{fig:DOS_Sn6O8_apart} (b)
shows the contribution of the Sn atoms which are still in between two O layers, i.e. the SnO$_2$-like Sn atoms. The valence band is mainly composed of SnO-like Sn atoms,
confirming our previous result that the valence band of the intermediate structures is similar to the valence band of SnO.

The conduction band of Sn$_3$O$_4$ is a hybridization of SnO-like Sn-p, O-p and SnO$_2$-like Sn-s states. When the O concentration increases, the conduction band minimum
decreases - leading to a smaller band gap, and these parabolic-shaped bands mainly consist of SnO$_2$-like Sn-s and O-p states, which are also the main contribution in the
conduction band of SnO$_2$, which is shown in Fig. \ref{fig:DOS_SnO2}.

Since both in conduction and valence band the contribution of SnO-like Sn-atoms is significant, a larger band gap for SnO$_2$ is reasonable, since it does not have SnO-like Sn atoms.
This is in line with the steep increase of the band gap for SnO$_2$, calculated with HSE, in Fig. \ref{fig:allgaps}.

\section{Conclusion}

In this study, we have investigated the stability, structural and electronic properties of Sn-O compounds with an O concentration between 50 \% (SnO) and 67 \% (SnO$_2$).
Seko et al. \cite{seko} already suggested a homologous series of (meta)stable compounds, with formula Sn$_{n+1}$O$_{2n}$, using ab initio calculations based on the PBE functional.
These structures have oxygen vacancies layered on (101) planes of the rutile lattice of SnO$_2$. In addition to this homologous series, we have found new (meta)stable structures,
arising from removing more than one O layer of SnO$_2$.

In addition to PBE calculations we also considered a vdW-corrected functional in order to better describe the Sn-Sn interlayer distances. In contrast to PBE the vdW functional
suggests the Sn$_{n+1}$O$_{2n}$ compounds to be metastable rather than stable. Besides PBE and the vdW-corrected functional, the HSE functional was used. We have presented for
the first time the electronic properties of the intermediate structures for which we observe a decrease of the band gap when (i) the O concentration increases and (ii) more O
layers are removed for a given concentration. Both the valence and the conduction band show a significant contribution of SnO-like Sn-atoms and when the O-concentration increases,
the SnO$_2$-like Sn-atoms start to dominate the conduction band.

Based on the results of this study we expect that other homologous series of Sn-O structures might be (meta)stable, namely by removing three or more (non-successive) O layers of
the SnO$_2$ superstructure, leading to a compound with a composition given by Sn$_{n+m}$O$_{2n}$ ($n = 2,3,..., m = 1,2, ...$ and $m < n$).

\section{Acknowledgments}
We gratefully acknowledge financial support from a GOA fund of the University of Antwerp. K.G. thanks the University of Antwerp for a PhD fellowship. The computational resources
and services used in this work were provided by the VSC (Flemish Supercomputer Center) and the HPC infrastructure of the University of Antwerp (CalcUA), both funded by the Hercules
Foundation and the Flemish Government - department EWI.

\bibliographystyle{unsrt}
\bibliography{sn-o}

\end{document}